\documentclass[sigconf]{acmart}
\usepackage[per-mode=symbol,per-symbol=p]{siunitx}
\usepackage[utf8]{inputenc}
\usepackage{adjustbox}
\usepackage{multirow}
\usepackage[ruled,vlined]{algorithm2e}
\usepackage{algorithmic}
\usepackage{graphicx}
\usepackage{textcomp}
\usepackage{xcolor}
\usepackage{balance}
\usepackage{enumitem}
\usepackage{setspace}
\usepackage{wrapfig}
\usepackage{listings}
\usepackage{color}
\usepackage{caption}
\usepackage{subcaption}
\usepackage{makecell}

\usepackage{amsmath}
\DeclareMathOperator*{\argmax}{arg\,max}
\DeclareMathOperator*{\argmin}{arg\,min}
\usepackage{mathtools}

\newcommand{\etal}{\emph{et al.}\xspace}
\newcommand{\ie}{\emph{i.e.}, }
\newcommand{\eg}{\emph{e.g.}, }
\newcommand{\etc}{\emph{etc. }}

\newcommand{\HAS}{\emph{HTTP Adaptive Streaming }}

\newcommand{\VVC}{\emph{Versatile Video Coding }}

\newcommand{\opte}{\texttt{OPTE}\xspace}

\newcommand{\scheme}{\texttt{QADRA}\xspace}

\newcommand{\EY}{$E_{\text{Y}}$}
\newcommand{\EU}{$E_{\text{U}}$}
\newcommand{\EV}{$E_{\text{V}}$}
\newcommand{\LY}{$L_{\text{Y}}$}
\newcommand{\LU}{$L_{\text{U}}$}
\newcommand{\LV}{$L_{\text{V}}$}
\newcommand{\h}{$h$}

\newcommand{\vig}[1]{\textcolor{black}{#1}}

\copyrightyear{2024}
\acmYear{2024}
\setcopyright{rightsretained}
\acmConference[MMSys '24]{ACM Multimedia Systems Conference 2024}{April 15--18, 2024}{Bari, Italy}
\acmBooktitle{ACM Multimedia Systems Conference 2024 (MMSys '24), April 15--18, 2024, Bari, Italy}\acmDOI{10.1145/3625468.3652172}
\acmISBN{979-8-4007-0412-3/24/04}

\begin{document}

\title{Quality-Aware Dynamic Resolution Adaptation Framework for Adaptive Video Streaming}

\author{Amritha Premkumar}
\email{amritha.premkumar@ieee.org}
\orcid{0009-0006-1480-4984}
\affiliation{
  \institution{\small{Department of Computer Science}}
  \institution{Rheinland-Pfälzische Technische
  Universität}
  \city{Kaiserslautern}
  \country{Germany}
}

\author{Prajit T Rajendran}
\email{prajit.rajendran@ieee.org}
\orcid{0000-0002-8283-9891}
\affiliation{
  \institution{\small{CEA, List, F-91120 Palaiseau}}
  \institution{Université Paris-Saclay}
  \city{Paris}
  \country{France}
}

\author{Vignesh V Menon}
\email{vignesh.menon@hhi.fraunhofer.de}
\orcid{0000-0003-1454-6146}
\affiliation{
  \institution{\small{Video Communication and Applications Dept}}
  \institution{Fraunhofer HHI}
  \city{Berlin}
  \country{Germany}
}

\author{Adam Wieckowski}
\email{adam.wieckowski@hhi.fraunhofer.de}
\orcid{0000-0003-0490-5803}
\affiliation{
  \institution{\small{Video Communication and Applications Dept}}
  \institution{Fraunhofer HHI}
  \city{Berlin}
  \country{Germany}
}

\author{Benjamin Bross}
\email{benjamin.bross@hhi.fraunhofer.de}
\orcid{0000-0002-1608-3774}
\affiliation{
  \institution{\small{Video Communication and Applications Dept}}
  \institution{Fraunhofer HHI}
  \city{Berlin}
  \country{Germany}
}

\author{Detlev Marpe}
\email{detlev.marpe@hhi.fraunhofer.de}
\orcid{0000-0002-5391-3247}
\affiliation{
  \institution{\small{Video Communication and Applications Dept}}
  \institution{Fraunhofer HHI}
  \city{Berlin}
  \country{Germany}
}

\renewcommand{\shortauthors}{Amritha Premkumar,~\etal}

\begin{abstract}
Traditional per-title encoding schemes aim to optimize encoding resolutions to deliver the highest perceptual quality for each representation. XPSNR is observed to correlate better with the subjective quality of VVC-coded bitstreams. Towards this realization, we predict the average XPSNR of VVC-coded bitstreams using spatiotemporal complexity features of the video and the target encoding configuration using an XGBoost-based model. Based on the predicted XPSNR scores, we introduce a \underline{Q}uality-\underline{A}ware \underline{D}ynamic \underline{R}esolution \underline{A}daptation (\scheme) framework for adaptive video streaming applications, where we determine the convex-hull online. Furthermore, keeping the encoding and decoding times within an acceptable threshold is mandatory for smooth and energy-efficient streaming. Hence, \scheme determines the encoding resolution and quantization parameter (QP) for each target bitrate by maximizing XPSNR while constraining the maximum encoding and/ or decoding time below a threshold. \scheme implements a JND-based representation elimination algorithm to remove perceptually redundant representations from the bitrate ladder. \scheme is an open-source Python-based framework published under the GNU GPLv3 license.\\
\textbf{Github}: \href{https://github.com/PhoenixVideo/QADRA}{https://github.com/PhoenixVideo/QADRA}\\
\textbf{Online documentation}: \href{https://phoenixvideo.github.io/QADRA/}{https://phoenixvideo.github.io/QADRA/}\\

\end{abstract}




\maketitle

\section{Introduction}

\HAS~(HAS) has emerged as the predominant method for delivering video content across a spectrum of internet speeds and device types~\cite{stockhammer_dynamic_2011}. Its core concept involves segmenting video content and encoding each segment at various bitrates and resolutions, known as \textit{representations}, which are then stored on standard HTTP servers. These representations facilitate continuous adaptation of video delivery to the fluctuating network conditions and diverse device capabilities of clients~\cite{bentaleb_survey_2019}. Typically, online streaming applications employ a predefined bitrate ladder, such as the one found in \textit{HTTP Live Streaming} (HLS)~\cite{apple_inc_http_nodate}, to ensure smooth and efficient content delivery.

\begin{figure}[t]
\centering
\begin{subfigure}{0.48\columnwidth}
    \centering
    \includegraphics[clip,width=\textwidth]{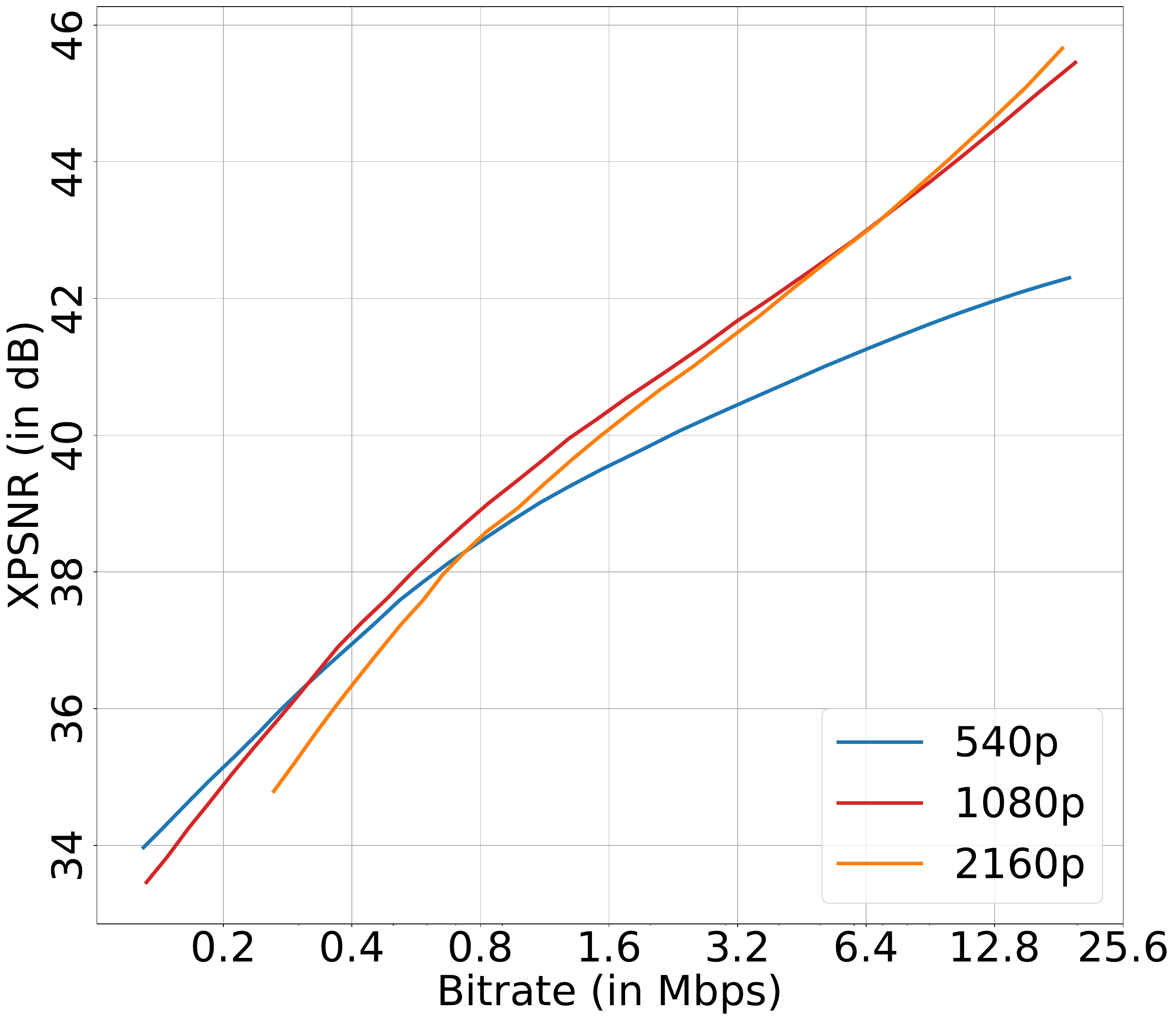}
    \includegraphics[clip,width=\textwidth]{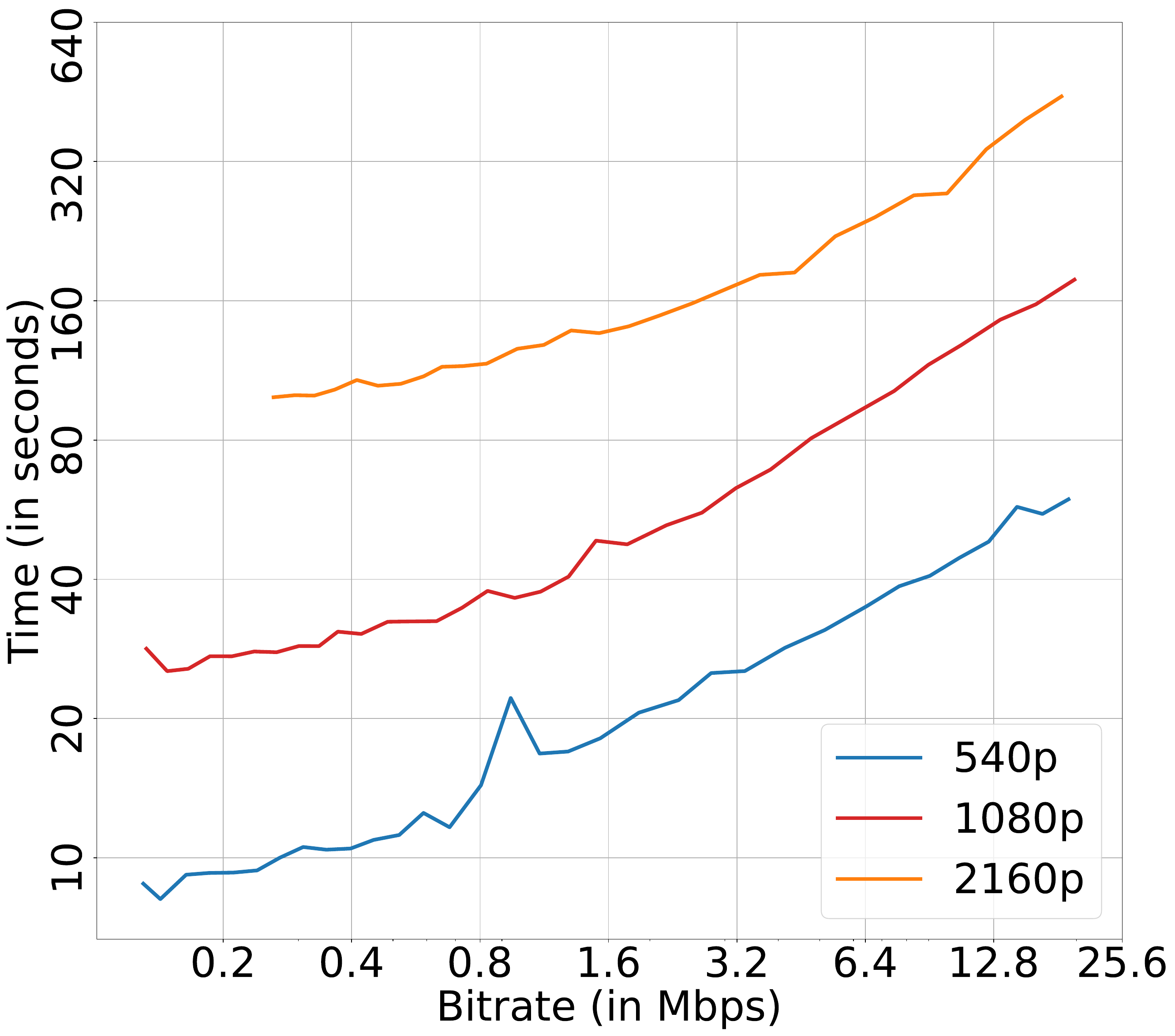}
    \caption{\textit{0100} ($E_{\text{Y}}$=22.40, $h$=4.70, $L_{\text{Y}}$=129.21)}    
    \label{fig:0100_intro}
\end{subfigure}
\begin{subfigure}{0.48\columnwidth}
    \centering
    \includegraphics[clip,width=\textwidth]{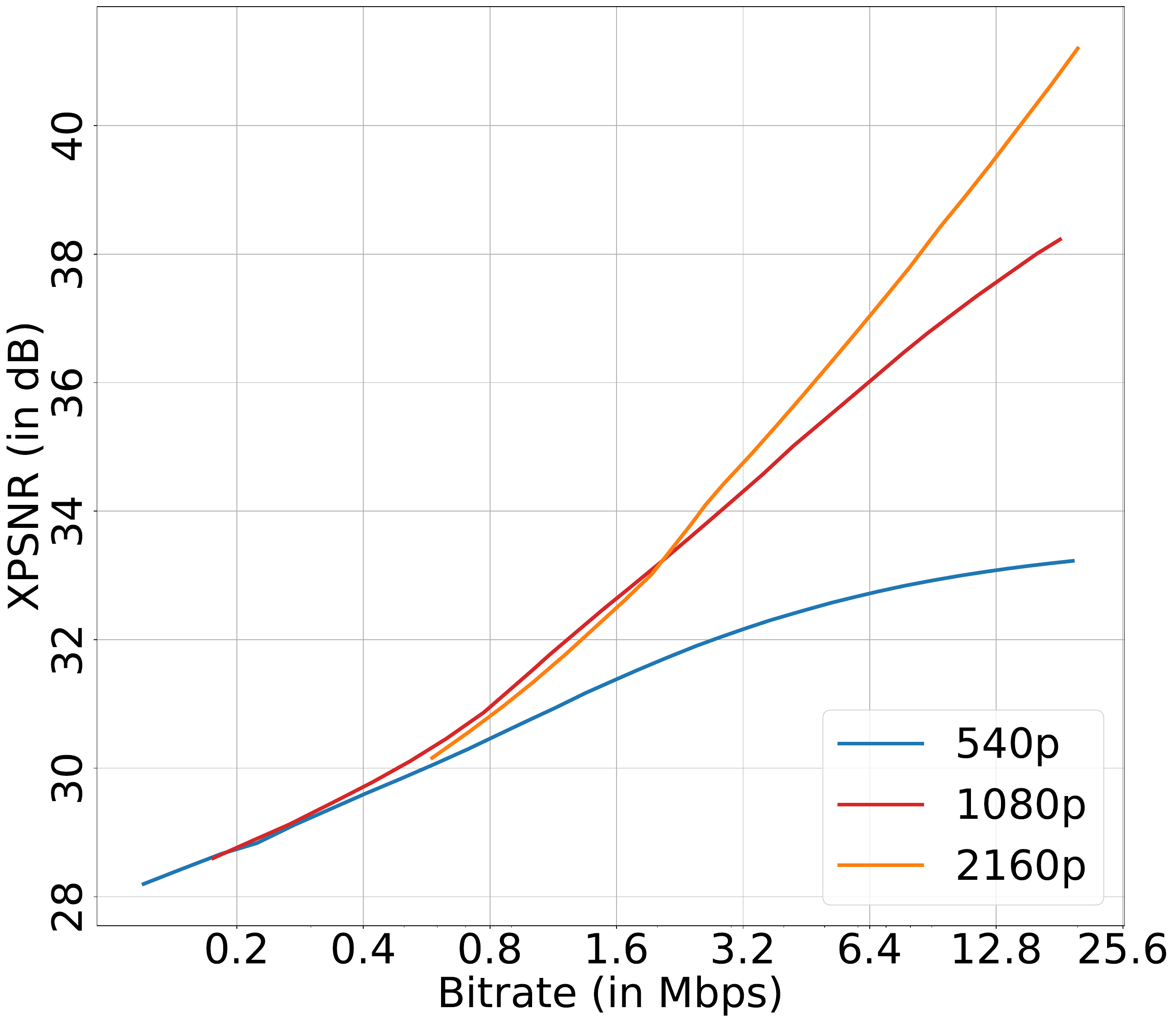}
    \includegraphics[clip,width=\textwidth]{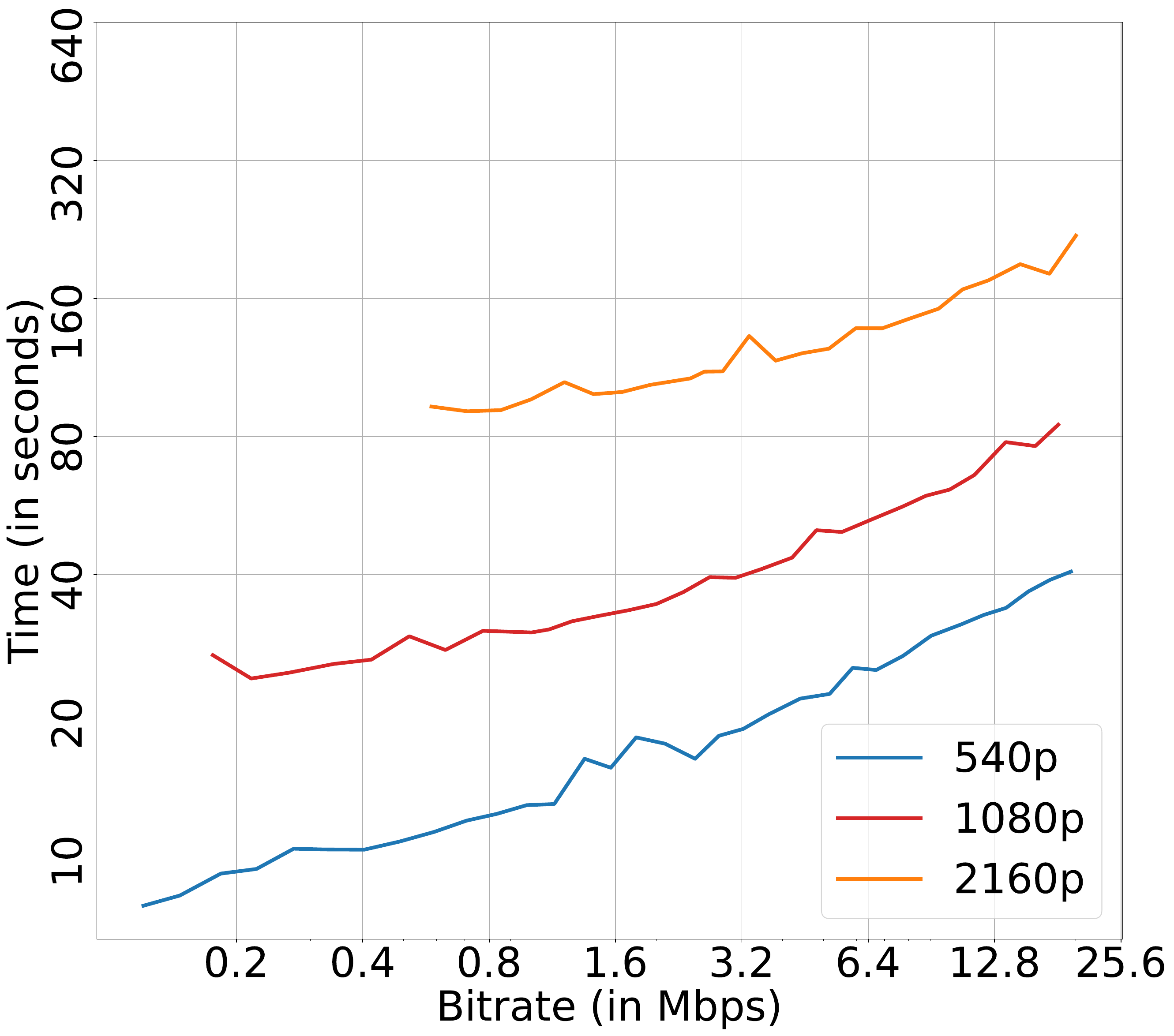}
    \caption{\textit{0250} ($E_{\text{Y}}$=20.40, $h$=8.72, $L_{\text{Y}}$=93.69)}    
    \label{fig:dolls_intro}
\end{subfigure}
\caption{Rate-distortion (RD) and rate-encoding time curves of representative sequences (segments) of Inter-4K dataset~\cite{stergiou_adapool_2023} encoded at 360p, 720p, 1080p and 2160p resolutions using VVenC encoder~\cite{wieckowski_vvenc_2021} at \textit{faster} preset. Here, XPSNR~\cite{helmrich_xpsnr_2020} is used as the quality metric.}
\label{fig:intro_convexhull}
\end{figure}

There is a growing interest in per-title encoding techniques aimed at improving the perceived quality of streamed content~\cite{de_cock_complexity-based_2016}. This innovative approach dynamically adjusts encoding parameters such as resolution~\cite{de_cock_complexity-based_2016, katsenou_content-gnostic_2019, bhat_case_2020, menon_jtps_2024}, framerate~\cite{menon_content_adaptive_2023}, and presets~\cite{menon_optimal_2024}, among others, based on content complexity and viewer preferences to optimize visual fidelity. Among these methods, dynamic resolution encoding has been extensively researched in adaptive streaming applications, where encoding resolutions are adjusted dynamically to maximize video quality~\cite{nasiri_ensemble_2022, telili_benchmarking_2022}. This approach ensures that perceptual quality remains high in visually intricate segments while conserving bandwidth by lowering resolution in less complex scenes. As demonstrated by rate-distortion plots in Figure~\ref{fig:intro_convexhull}, the optimal resolution, which maximizes perceptual quality (measured in terms of XPSNR~\cite{helmrich_xpsnr_2020}), varies depending on content complexity. By adapting resolution per segment, the streaming system efficiently allocates resources, prioritizing high-quality representation where it's most beneficial. Ultimately, dynamic resolution per-title encoding aims to balance perceptual quality and bandwidth efficiency, delivering an immersive and captivating streaming experience~\cite{de_cock_complexity-based_2016}.

Reducing encoding time is critical in streaming applications since it contributes to environmental sustainability. Encoding processes in data centers require substantial computational resources and energy consumption, especially with new codecs such as \VVC~(VVC)~\cite{bross_overview_2021, kaafarani_evaluation_2021}. The streaming industry can reduce its carbon footprint and energy consumption by lowering encoding time~\cite{menon_energy_efficient_2024,herglotz_estimating_2015}. Furthermore, reducing decoding time on the client side reduces stall events and buffering time, contributing to a smooth viewing experience~\cite{farhat_energy_2023,azimi_lable_2024}. 

In this paper, the main contributions are as follows:
\begin{enumerate}
\item A quality-aware encoding resolution selection framework to maximize the perceptual quality (in terms of XPSNR) of video segments based on their spatiotemporal complexity, target bitrate, and the encoding and/or decoding time constraint for VVC-based streaming environments.
\item Comprehensive analysis of the proposed framework for various encoding time thresholds regarding compression efficiency and encoding latency.
\end{enumerate}

\textit{\textbf{Outline: }}
The remainder of this paper is organized as follows. Section~\ref{sec:dyn_res} discusses the related work on dynamic resolution encoding in the context of adaptive video streaming. The proposed \scheme framework is explained in Section~\ref{sec:prop_framework}, while Section~\ref{sec:results} presents the experimental results. Finally, Section~\ref{sec:conclusion} concludes the paper.

\section{Related work}
\label{sec:dyn_res}
Most state-of-the-art dynamic resolution per-title encoding methods are based on choosing a particular resolution that provides better visual quality for a given bitrate range. 

Katsenou~\etal~\cite{katsenou_content-gnostic_2019} uses machine learning to identify the most effective bitrate range for each resolution. The method extracts spatiotemporal features and statistics from sequences at their original resolution. Then, it employs machine learning methods to predict the quantization parameters (QPs) at which the rate-distortion curves across the different resolutions intersect. $(\Tilde{r}-1)\times 2$ encodes must be performed to determine the bitrates at which resolutions should be switched.  This content-gnostic approach has been claimed to reduce the number of encodings required compared to other methods (by 81\% - 94\%) compared to the bruteforce encoding approach. It uses constant quantization parameter (CQP) encodes, which are not used in real-time streaming applications. 
Another method proposed by Bhat~\etal~\cite{bhat_case_2020} uses machine learning to predict the resolution without requiring multiple encodings. Features from the low-resolution encoding of the first few frames are input to a random forest model to predict better-performing resolution for a decision period. Similarly, Zabrovskiy~\etal~\cite{zabrovskiy_faust_2021} used an artificial neural network to predict an optimized bitrate ladder for each scene, optimized based on the YPSNR quality metric. These methods produce \textit{latency} significantly higher than the accepted latency in live streaming. 
\opte~\cite{menon_jtps_2024} uses random forest models to predict optimized resolution, yielding the highest VMAF~\cite{blog_vmaf_2018} using spatiotemporal features extracted for each segment. However, \opte does not consider encoding latency constraint during the optimized resolution prediction. 

To summarize, current related work lacks encoding latency constraints while selecting the optimized encoding resolution, and most state-of-the-art methods need pre-encodings that yield significant latency and energy consumption. Furthermore, discussion on dynamic resolution per-title encoding for VVC-based streaming platforms is limited.

\section{Quality-aware dynamic resolution adaptation (\scheme)}
\label{sec:prop_framework}
\begin{figure*}[t]
\centering
\includegraphics[width=0.9\linewidth]{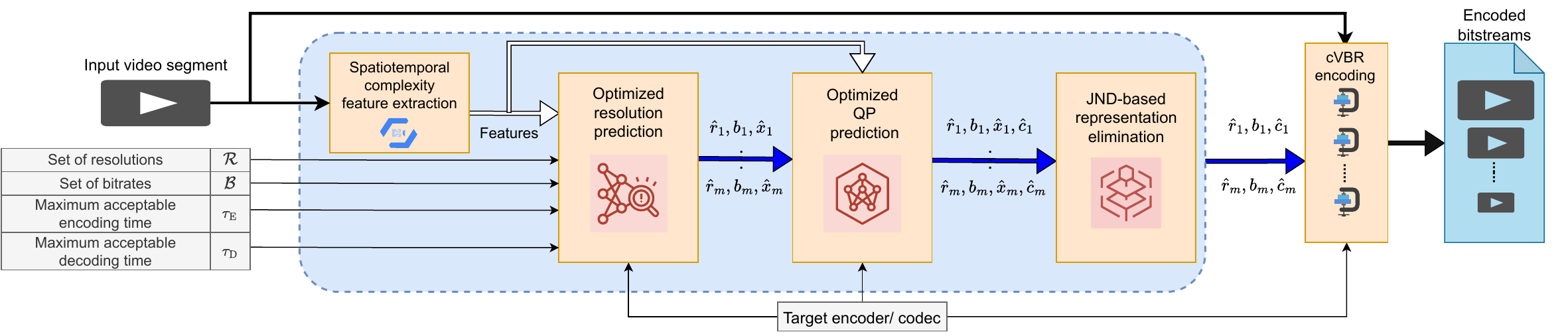}
\vspace{-0.63em}
\caption{Encoding using \scheme framework envisioned in this paper.}
\vspace{-0.43em}
\label{fig:contribution}
\end{figure*}

Striking the right balance between offering high-quality, high-resolution streams and minimizing encoding and/or decoding time and energy consumption is crucial for adaptive streaming platforms to ensure responsive and uninterrupted playback experiences across various end-user devices and network environments. In line with this perspective, this paper proposes a latency-aware dynamic encoding resolution encoding scheme (\scheme) to maximize the perceived quality of video segments based on the video content complexity, target bitrate, and the encoding time constraint. As shown in Figure~\ref{fig:contribution}, \scheme is classified into four steps:
\begin{enumerate}
    \item spatiotemporal complexity feature extraction (Section~\ref{sec:features}),
    \item optimized resolution prediction (Section~\ref{sec:res_pred}),
    \item optimized QP prediction (Section~\ref{sec:crf_pred}) ,
    \item JND-based representation elimination (Section~\ref{sec:jnd_elim}),    
\end{enumerate}

\subsection{Spatiotemporal complexity feature extraction}
\label{sec:features}
This process involves analyzing the video content in both spatial and temporal dimensions, capturing essential information about object movements, scene changes, and visual details. Prediction models can comprehensively understand the content complexity and characteristics by extracting relevant spatiotemporal features, such as motion vectors, texture patterns, and frame-to-frame differences. \scheme uses seven DCT-energy-based features extracted by Video Complexity Analyzer (VCA) v2.0~\cite{menon_green_2023}: the average texture energy (\EY), the average gradient of the luma texture energy (\h), the average luminescence (\LY), Average chroma texture energy of U and V channels (\EU~and \EV), and the average chrominescence of U and V channels (\LU~and \LV) as the content complexity features of video segments. The features of the videos of the Inter-4K dataset~\cite{stergiou_adapool_2023} are compiled in the \texttt{/Dataset/} folder of the repository.

\subsection{Optimized resolution prediction}
\label{sec:res_pred}
The objective of selecting the optimized resolution based on bitrate and video complexity features is decomposed into two parts:
\begin{enumerate}
    \item designing models to predict the encoding and/ or decoding time and the perceptual quality;
    \item developing a function to obtain the optimized resolution based on the predicted encoding and/ or decoding times and perceptual quality for each available encoding resolution.
\end{enumerate}

\textit{\textbf{Modeling: }}  
The perceptual quality (in terms of XPSNR) $x_{(r_{t},b_{t})}$, encoding time $e_{(r_{t},b_{t})}$, and decoding time $d_{(r_{t},b_{t})}$ of the representation $(r_{t},b_{t})$ relies on video complexity features \{\EY, \h, \LY, \EU, \EV, \LU, \LV\}, encoding resolution $r_t$, and target bitrate $b_t$ parameters:
\begin{align}
\label{eq:v_pred}
    x_{\left(r_{t},b_{t}\right)} &= f_{\text{V}}\left(E_{\text{Y}}, h, L_{\text{Y}}, E_{\text{U}},E_{\text{V}}, L_{\text{U}}, L_{\text{V}}, r_{t}, b_{t}\right);\\
\label{eq:s_pred}
    e_{\left(r_{t},b_{t}\right)} &= f_{e}\left(E_{\text{Y}}, h, L_{\text{Y}}, E_{\text{U}},E_{\text{V}}, L_{\text{U}}, L_{\text{V}}, r_{t}, b_{t}\right);\\
\label{eq:d_pred}
    d_{\left(r_{t},b_{t}\right)} &= f_{d}\left(E_{\text{Y}}, h, L_{\text{Y}}, E_{\text{U}},E_{\text{V}}, L_{\text{U}}, L_{\text{V}}, r_{t}, b_{t}\right).    
\end{align}

Spatio-temporal features encapsulate intricate spatial details and temporal dynamics within the video segment and help assess the video fidelity~\cite{cai_real-time_2019}. Including resolution, bitrate, framerate, and preset parameters in the models acknowledges the interplay between compression efficiency, temporal smoothness, and spatial clarity in shaping perceived quality~\cite{menon_transcoding_2023}. A higher resolution, or bitrate, may improve the quality and increase the file size of the video segment. Similarly, a higher resolution, or bitrate, can reduce the encoding and decoding speed. Notably, encoding and decoding speeds largely depend on hardware-level parameters like RAM capacity, CPU threads, \etc~\cite{menon_content_adaptive_2023}. 

\textit{\textbf{Optimization: }} 
\scheme optimizes the perceptual quality of encoded video segments while adhering to real-time processing constraints. It predicts the optimized resolution of the $t$\textsuperscript{th} representation to maximize the compression efficiency while maintaining the encoding time below the threshold  $\tau_{\text{L}}$. The optimization function is:
\begin{align}
   \hat{r}_{t} &= \argmax_{r \in \mathcal{R}} \hat{x}_{(r,b_{t})} & c.t. \hspace{1em} \hat{e}_{(r,b_{t})} \leq \tau_{\text{E}}, \hspace{0.5em}\hat{d}_{(r,b_{t})} \leq \tau_{\text{D}}.  
\end{align}
where $\hat{x}_{(r,b_{t})}$, $\hat{e}_{(r,b_{t})}$, and $\hat{d}_{(r,b_{t})}$ are the predicted XPSNR, encoding and decoding speeds of the representation $(r,b_{t})$.

\textit{\textbf{Implementation: }}
The \texttt{select\_best\_resolution} method selects the best resolution from predefined resolutions based on the predicted encoding time and the target bitrate. It aims to find the resolution that maximizes the XPSNR within the target encoding time. Firstly, the variables to store predicted XPSNR values (\texttt{xpsnr}) and predicted encoding times (\texttt{time}) are initialized for each resolution in the predefined list. For each resolution in the list, the XPSNR (\texttt{xpsnr}) and encoding time (\texttt{time}) are predicted using the provided features and bitrate. These predictions are made using separate methods (\texttt{predict\_xpsnr} and \texttt{predict\_enc\_time}). 
The highest predicted XPSNR value (\texttt{highest\_xpsnr}) that satisfies the target encoding time constraint (\texttt{tl}) is identified. If a resolution yields the highest XPSNR within the target encoding time, that resolution is selected as the predicted resolution (\texttt{predicted\_resolution}). 
If no resolution satisfies the target encoding time constraint, the predicted resolution remains unchanged (defaulting to the first resolution in the list). The predicted resolution is adjusted based on the bitrate using the \texttt{get\_resolution\_based\_on\_bitrate} method. 

We employ a data-driven approach to select the resolution that maximizes video quality within the given encoding time constraint, ultimately optimizing the encoding process for the specified bitrate.

\subsection{Optimized QP prediction}
\label{sec:crf_pred} 
Predicting the QP helps ensure consistent video quality throughout the stream. It allows the encoder to allocate bits judiciously, preventing underallocation (resulting in poor quality) or over-allocation (wasting bandwidth) of bits for encoding.

\textit{\textbf{Modeling: }} 
The QP $q_{(r_{t},b_{t})}$ relies on video complexity features \{\EY, \h, \LY, \EU, \EV, \LU, \LV\}, encoding resolution $r_t$, and target bitrate $b_t$ parameters:
\begin{align}
\label{eq:c_pred}
    q_{\left(r_{t},b_{t}\right)} &= f_{\text{Q}}\left(E_{\text{Y}}, h, L_{\text{Y}}, E_{\text{U}},E_{\text{V}}, L_{\text{U}}, L_{\text{V}}, r_{t}, b_{t}\right).
\end{align}
Content with intricate details, textures, or sharp edges demands a lower QP to represent these features accurately in the encoded video. Similarly, segments with fast motion, frequent scene changes, or dynamic content require a lower QP to capture the rapid changes between frames accurately~\cite{menon_all_intra_2023}. 

\textit{\textbf{Optimization: }}
The mathematical formulation of the QP optimization to yield a bitrate as close to the target bitrate as possible can be expressed as follows: 
\begin{align}
   \hat{q}_{(r,b_{t})} &= \argmin_{q \in [q_{\text{min}},q_{\text{max}}]} \mid \hat{b}_{(r,c)}- b_{t} \mid .  
\end{align}
A loss function measures the deviation between the target and predicted bitrate. The objective is to find the QP that minimizes the loss function.

\textit{\textbf{Implementation: }}
\vig{The \texttt{predict\_qp} function takes input features extracted from the video segment, along with the resolution (normalized to a range of [0, 1]) and bitrate. These inputs are concatenated into a feature vector. The function retrieves two pre-trained machine learning models: one for predicting the minimum QP (\texttt{min\_model}) and another for predicting the maximum QP (\texttt{max\_model}). The resolution (normalized) is appended to the feature vector. The feature vector is passed to the minimum and maximum QP prediction models to obtain predicted QP values (\texttt{b1} and \texttt{b2}). 
The function uses linear interpolation to compute a predicted QP value (\texttt{qp\_pred}) based on the bitrate. The predicted QP is calculated based on the equation of a line passing through two points: (x1, b1) and (x2, b2), where x1 and x2 are predefined values (10 and 50) and b1 and b2 are the predicted QP values corresponding to these points. The predicted QP value is constrained from 10 to 50 in our implementation. The function returns the predicted QP value (\texttt{qp\_pred}) as an integer.}

\subsection{JND-based representation elimination}
\label{sec:jnd_elim}
\scheme~uses the JND-based representation elimination algorithm proposed in our previous work~\cite{menon_content_adaptive_2023}. To avoid the perceptual redundancy of the bitrate ladder, if the predicted quality difference between two representations is lower than the JND~\cite{zhu2022framework}, the higher bitrate representation amongst them is eliminated. Furthermore, when the predicted quality is greater than $v_{\text{T}}$, \ie the threshold above which the representation is deemed perceptually lossless, the corresponding representation is eliminated from the bitrate ladder~\cite{menon_mcps_2023}.  

\textit{\textbf{\vig{Implementation:} }} 
\vig{The \texttt{jnd\_elimination} function is implemented to eliminate representations based on the JND criterion. The \texttt{jnd\_elimination} function takes a list of representations with JND features as input and returns a subset of representations based on the JND criterion.}
\begin{enumerate}
    \item Initialize an empty list of representations to store the selected representations.
    \item If the JND threshold (\texttt{self.jnd}) is set to 0, return the input list of representations.
    \item Otherwise, iterate through the list of representations.
    \item Add the first representation to the representations list.
    \item If the XPSNR value of the current representation exceeds the maximum XPSNR threshold (\texttt{self.max\_xpsnr}), return the representations list.
    \item Iterate through the remaining representations in the list.
    \item If the difference in XPSNR between the current representation and the last selected representation is greater than or equal to the JND threshold, add the current representation to the representations list.
    \item Return the representations list if the XPSNR value of the current representation exceeds the maximum XPSNR threshold.
    \item Return the final representations list.
\end{enumerate}

\subsection{Commandline options}
\label{sec:cli_options}
The following command line options are included in the prototype:
\begin{enumerate}
    \item \emph{maxEncTime}: This option allows the user to specify the maximum acceptable time for encoding each representation. It is a crucial parameter for controlling the encoding resolutions allocated for each representation. Default: 9999.  
    \item \emph{maxDecTime}: This option allows the user to specify the maximum acceptable time for decoding each representation on the client side (if available). Default: 9999.    
    \item \emph{codec}: This option lets the user specify the codec (encoder-decoder configuration) used. Users can choose from popular codecs depending on their specific video encoding or decoding requirements. Default: vvenc.
    \item \emph{resultCsv}: This option takes the path to the optimized bitrate ladder comma-separated values (CSV) file. This CSV file likely contains optimized encoding resolution values for efficient video encoding and delivery. Default: results.csv.
    \item \emph{rmax}: This option represents the maximum supported resolution. Users can set this option to define the highest resolution permissible for the video encoding task. It ensures that encoding operates within the specified resolution limits. Default: 2160.
    \item \emph{maxQuality}: This option allows users to set the maximum acceptable XPSNR score. Setting a maximum threshold helps ensure that only videos are encoded below a certain quality level. This is especially used to eliminate perceptually lossless representations. Default: 100.
    \item \emph{jnd}: This option allows users to set the threshold for perceptual differences (in terms of XPSNR), ensuring that the process considers only noticeable differences in video quality. Default: 0.
\end{enumerate}
These CLI options provide a range of customization for users, enabling them to tailor the prototype's behavior according to their specific needs and preferences in video streaming.

\subsection{Prediction models}
80 \% of the videos in the Inter-4K Dataset~\cite{stergiou_adapool_2023} is used to train the prediction models. Encodings are run on a dual-processor server with Intel Xeon Gold 5218R (80 cores, frequency at 2.10 GHz), where each encoding instance uses four CPU threads. The sequences are encoded at 60\,fps using VVenC v1.10~\cite{wieckowski_vvenc_2021} using preset 0 (\textit{faster}), and QPs ranging from $q_{\text{min}}$ to $q_{\text{max}}$.  The spatiotemporal features, \{\EY, \h, \LY, \EU, \EV, \LU, \LV\} are extracted using VCA v2.0~\cite{menon_green_2023} running as a pre-processor using four CPU threads with multi-threading and x86 SIMD optimizations. The procedure to generate the dataset for training is illustrated in Algorithm~\ref{algo:dataset_gen}.

\begin{algorithm}[t]
\caption{Training dataset generation.}
\footnotesize
\textbf{Inputs:}\\
\quad $\mathcal{R}$: set of supported resolutions \\
\quad $q_{\text{min}}$: minimum QP \\
\quad $q_{\text{max}}$: maximum QP \\

\For{each training video segment}{
    Run VCA and get \{\EY, \h, \LY, \EU, \EV, \LU, \LV \} \\
    \For{each $r \in \mathcal{R}$}{
        \For{each $q \in [q_{\text{min}}, q_{\text{max}}]$}{
            Encode segment with QP $q$ \;
            Record \EY, \h, \LY, \EU, \EV, \LU, \LV, $r$, $q$, achieved bitrate $b'$, XPSNR $x'$, and PSNR $p'$ \;
        }
    }
}
\label{algo:dataset_gen}
\end{algorithm}

We trained the XPSNR prediction models using multiple regressors, including extra-trees, XGBoost~\cite{chen_xgboost_2016}, and random forests~\cite{breiman_random_2001} and observed that the XGBoost regressor performed the best consistently using our feature set. A grid search is performed to explore different combinations of hyperparameter values, and we selected \emph{max\_depth}=10, and \emph{n\_estimators}=400 that maximized performance. 
Our predictive modeling framework employs a cascading approach to predict both bitrate and encoding time for a given QP. This method involves training distinct models for minimum and maximum QP values ($q_{\text{min}}$ and $q_{\text{max}}$, respectively), enabling the prediction of both the maximum and minimum bitrate, as well as the maximum and minimum encoding time. The linear relationship between QP and the logarithm of time and between QP and the logarithm of bitrate underpins the success of this approach. Since the points ($q_{\text{min}}$, $b_{\text{max}}$)  and ($q_{\text{max}}$, $b_{\text{min}}$) are estimated, the optimized QP for a target bitrate $b$ is determined using linear regression. Similarly, using the estimations from the encoding time models, we obtain the points ($q_{\text{min}}$, $\hat{t}_{\text{max}}$) and ($q_{\text{max}}$, $\hat{t}_{\text{min}}$) from which the corresponding encoding time for the QP value is determined.

\vig{We have evaluated their performance across various video content types, resolutions, and bitrates, demonstrating consistent and reliable predictions. Additionally, rigorous cross-validation techniques have been employed to assess model performance and ensure its robustness against overfitting. Despite our confidence in the models, we acknowledge potential scenarios where they might underperform or fail to predict encoding parameters accurately. Some factors that could contribute to underperformance include out-of-distribution data and extreme conditions such as highly complex content. As encoding techniques evolve and new video formats emerge, the models may require periodic updates to maintain their predictive accuracy and generalizability. }

\subsection{\vig{Scalability and adaptability}}
\vig{The modular design of \scheme allows easy integration with existing streaming infrastructure and workflow automation systems, facilitating deployment across large-scale streaming platforms. Its predictive models are built to handle a wide range of input data, enabling efficient processing of large datasets and rapid generation of encoding parameter recommendations. \scheme's architecture can be optimized for parallel processing, leveraging distributed computing resources to scale seamlessly with increasing demand. \scheme is adaptable to evolving technologies, including immersive media streaming formats such as virtual reality (VR) and augmented reality (AR). Its predictive models can be trained on datasets encompassing diverse video formats, resolutions, and encoding techniques, enabling it to adapt to emerging standards and formats in immersive media streaming.}


\begin{table}[t]
\caption{Experimental parameters used to evaluate \scheme.}
\centering
\resizebox{\columnwidth}{!}{
\begin{tabular}{l||c|c|c|c|c|c}
\specialrule{.12em}{.05em}{.05em}
\specialrule{.12em}{.05em}{.05em}
\emph{Parameter} & \multicolumn{6}{c}{\emph{Values}}\\
\specialrule{.12em}{.05em}{.05em}
\specialrule{.12em}{.05em}{.05em}
$\mathcal{R}$ & \multicolumn{6}{c}{\{ 360, 432, 540, 720, 1080, 1440, 2160 \} } \\
\hline
$\mathcal{B}$ & 0.145 & 0.300 & 0.600 & 0.900 & 1.600 & 2.400 \\
              & 3.400 & 4.500 & 5.800 & 8.100 & 11.600 & 16.800 \\
\hline
 $\tau_{\text{E}}$ & \multicolumn{6}{c}{ 100s, 200s, 400s, 800s, $\infty$ } \\
\hline
 $\tau_{\text{D}}$ & \multicolumn{6}{c}{$\infty$ } \\
\hline
 $r_{\text{max}}$ & \multicolumn{2}{c|}{720} & \multicolumn{2}{c|}{1080} & \multicolumn{2}{c}{2160} \\
\hline
Target encoder & \multicolumn{6}{c}{VVenC (\textit{faster} preset)} \\
\hline
Target decoder & \multicolumn{6}{c}{VTM} \\
\hline
CPU threads & \multicolumn{6}{c}{4} \\
\specialrule{.12em}{.05em}{.05em}
\specialrule{.12em}{.05em}{.05em}
\end{tabular}
}
\label{tab:exp_par}
\end{table}

\begin{figure}[t]
\centering
\begin{subfigure}{0.448\columnwidth}
    \centering
    \includegraphics[width=0.94\textwidth]{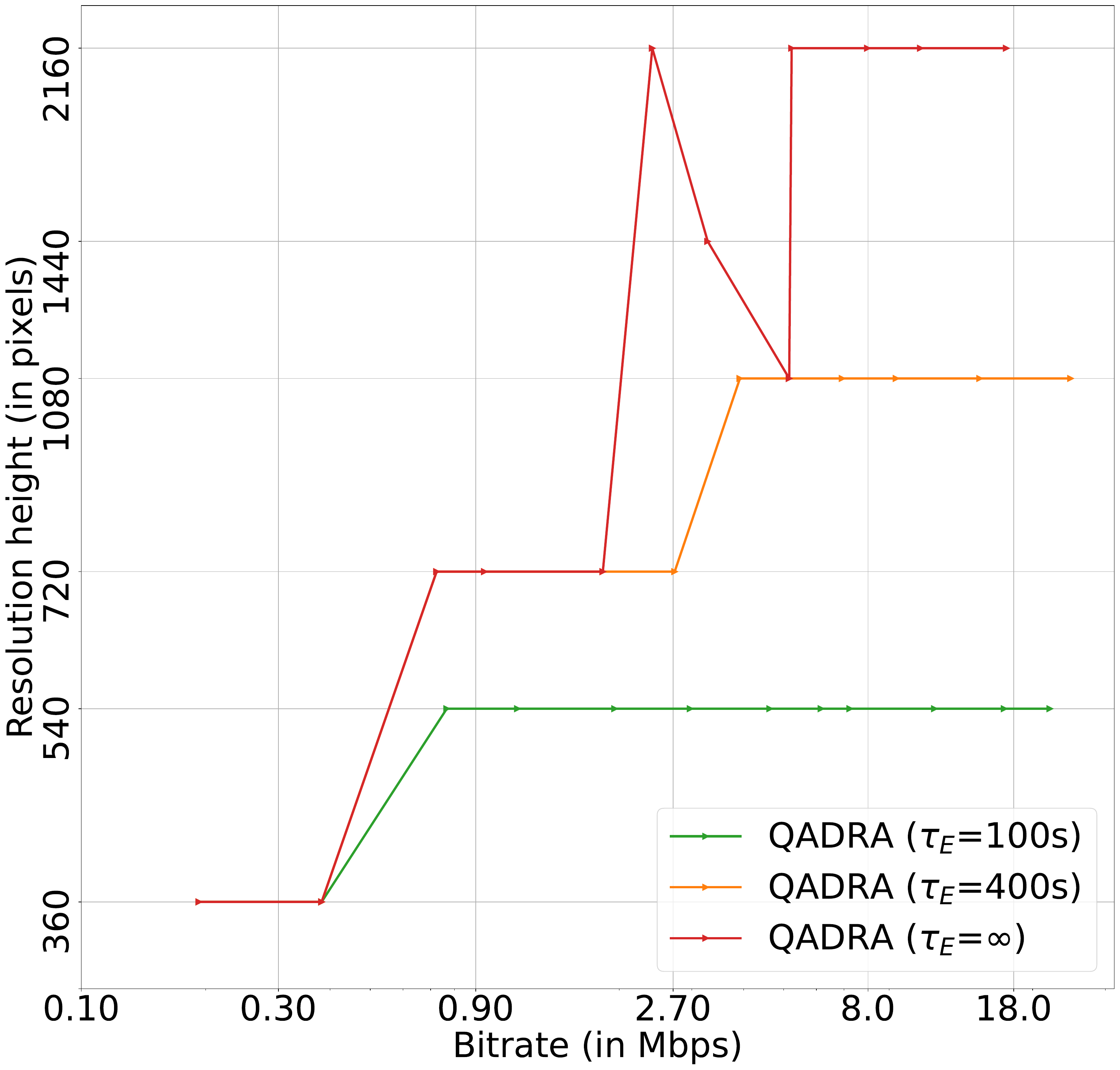}
    \caption{\textit{0001}}
\end{subfigure}
\hfill
\begin{subfigure}{0.448\columnwidth}
    \centering    
    \includegraphics[width=0.94\textwidth]{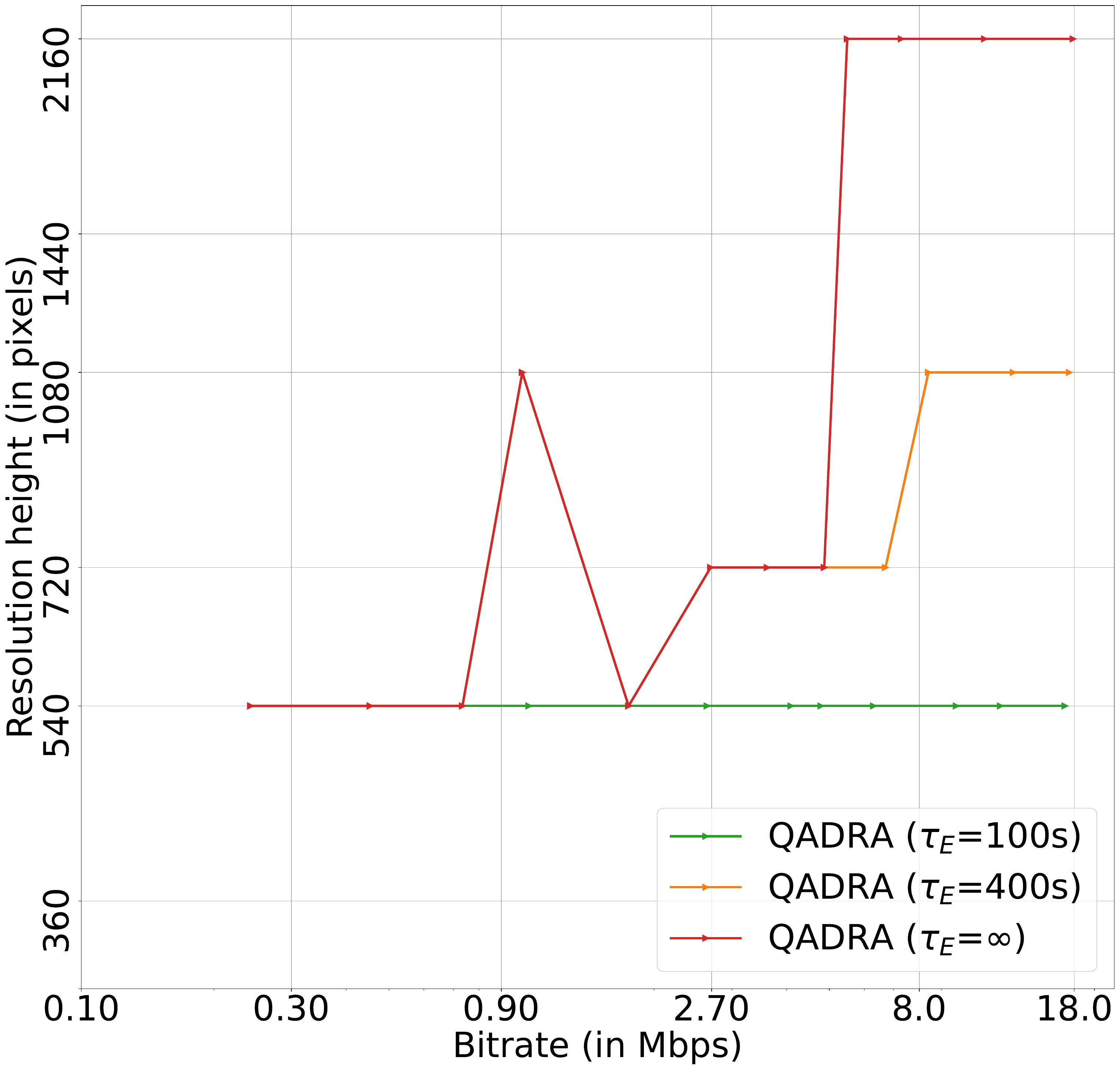}
    \caption{\textit{0334}}
\end{subfigure}
\begin{subfigure}{0.448\columnwidth}
    \centering    
    \includegraphics[width=0.94\textwidth]{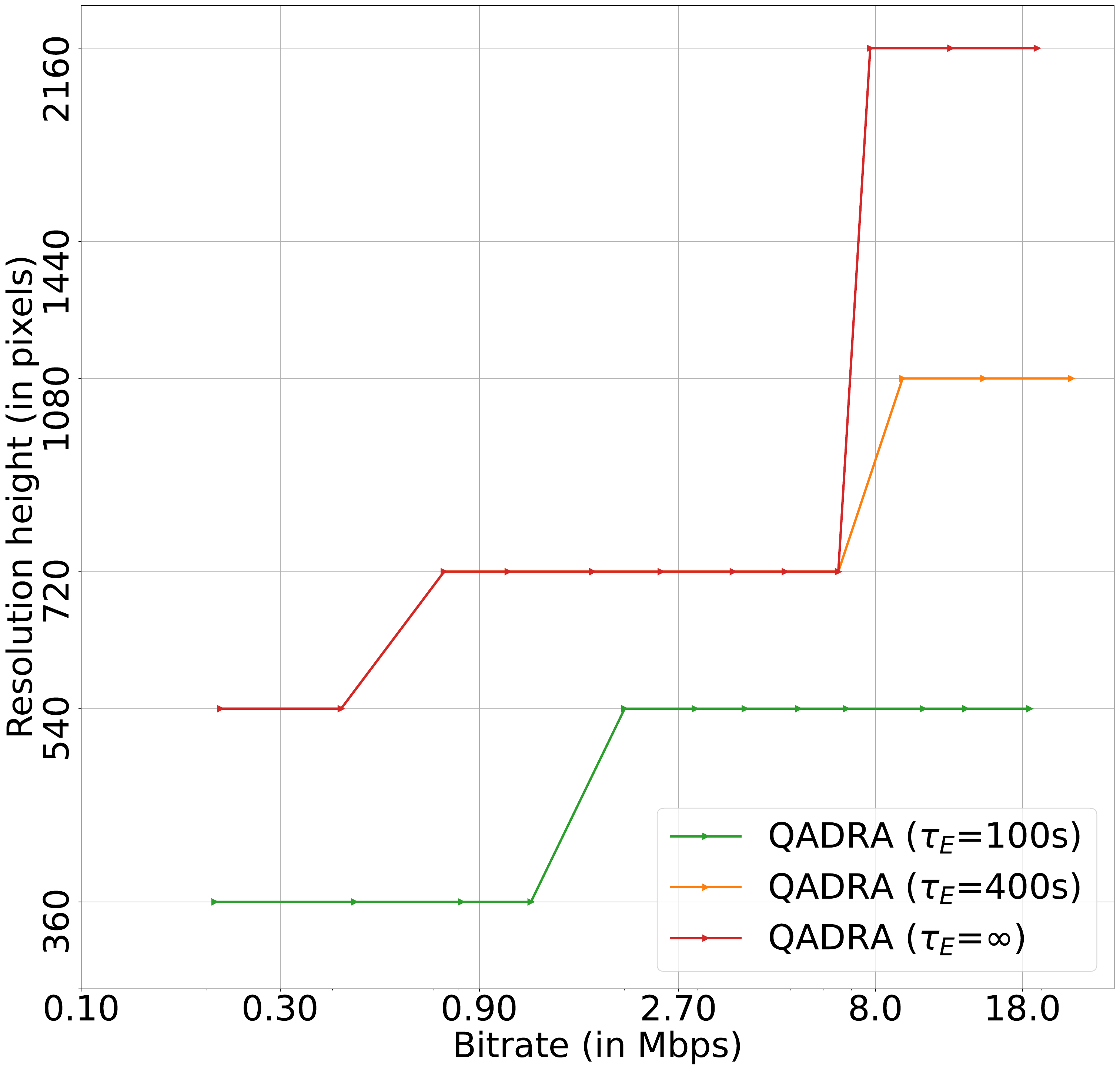}
    \caption{\textit{0412}}
\end{subfigure}
\hfill
\begin{subfigure}{0.448\columnwidth}
    \centering
    \includegraphics[width=0.99\textwidth]{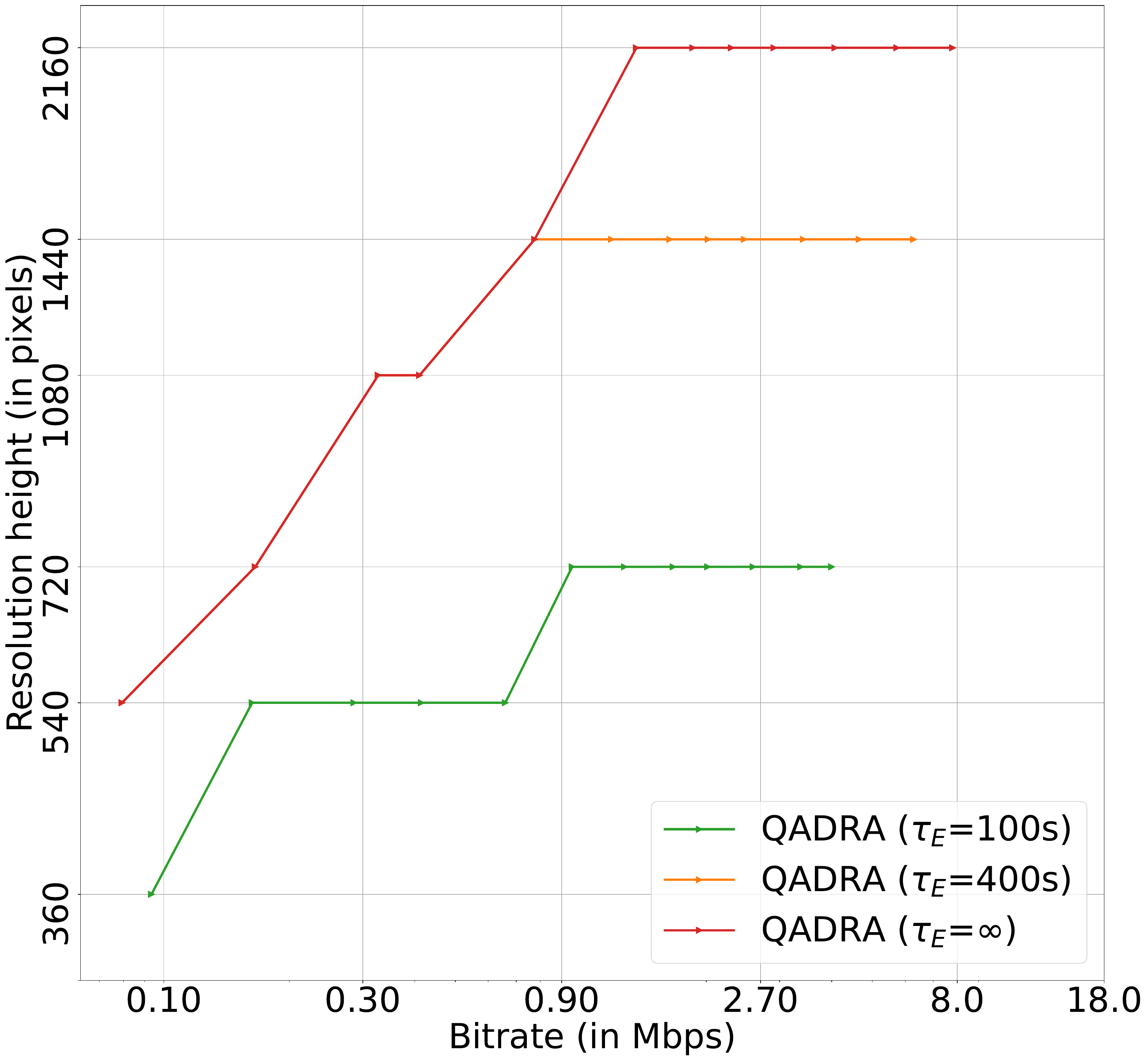}
    \caption{\textit{0744}}
\end{subfigure}
\caption{Selected encoding resolutions of representative video sequences (segments) using \scheme ($r_{\text{max}}=2160$).}
\label{fig:res_res}
\end{figure}

\begin{figure}[t]
\centering
\begin{subfigure}{0.223\textwidth}
    \centering
    \includegraphics[clip,width=0.9\textwidth]{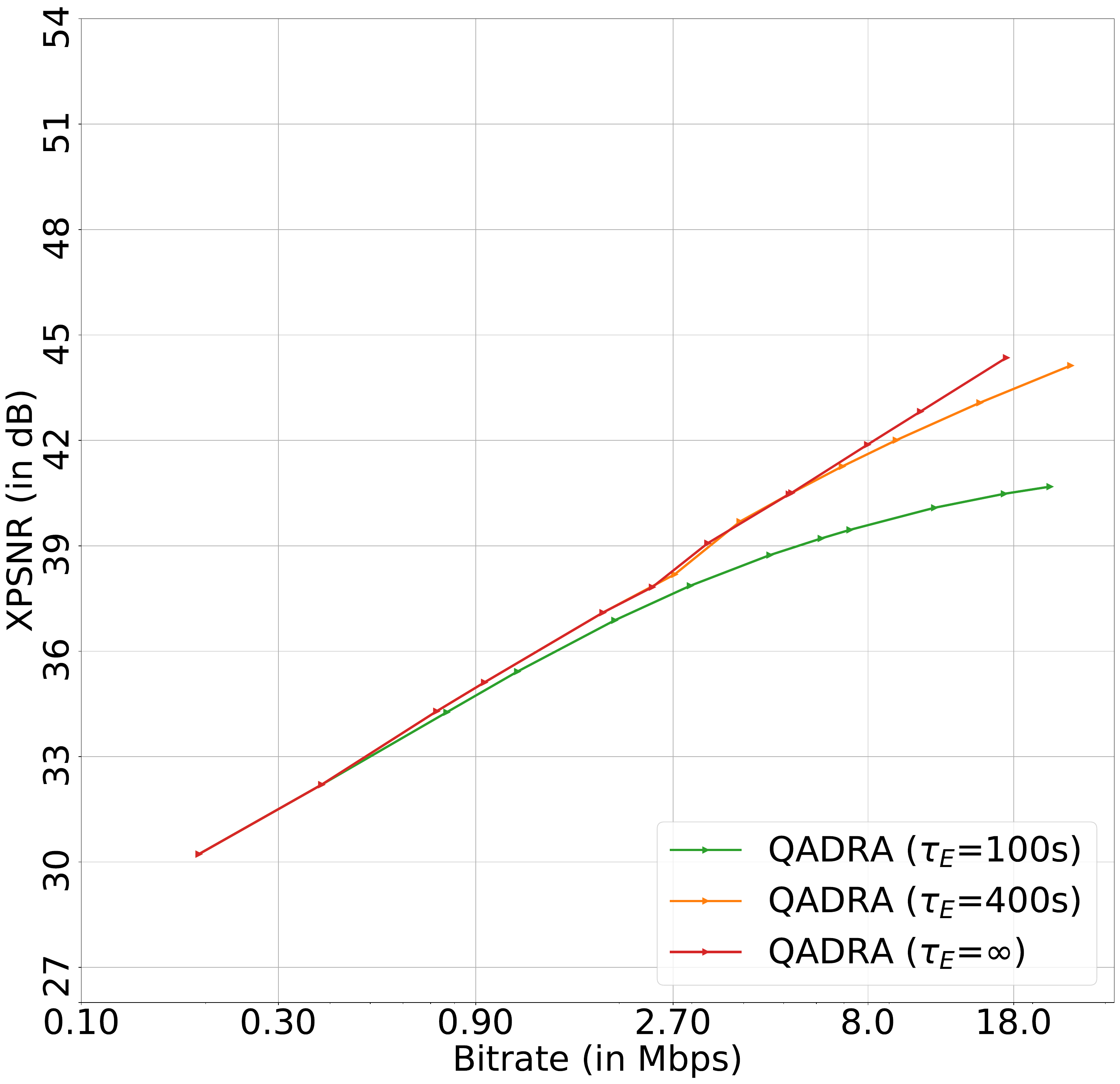}  
    \includegraphics[clip,width=0.9\textwidth]{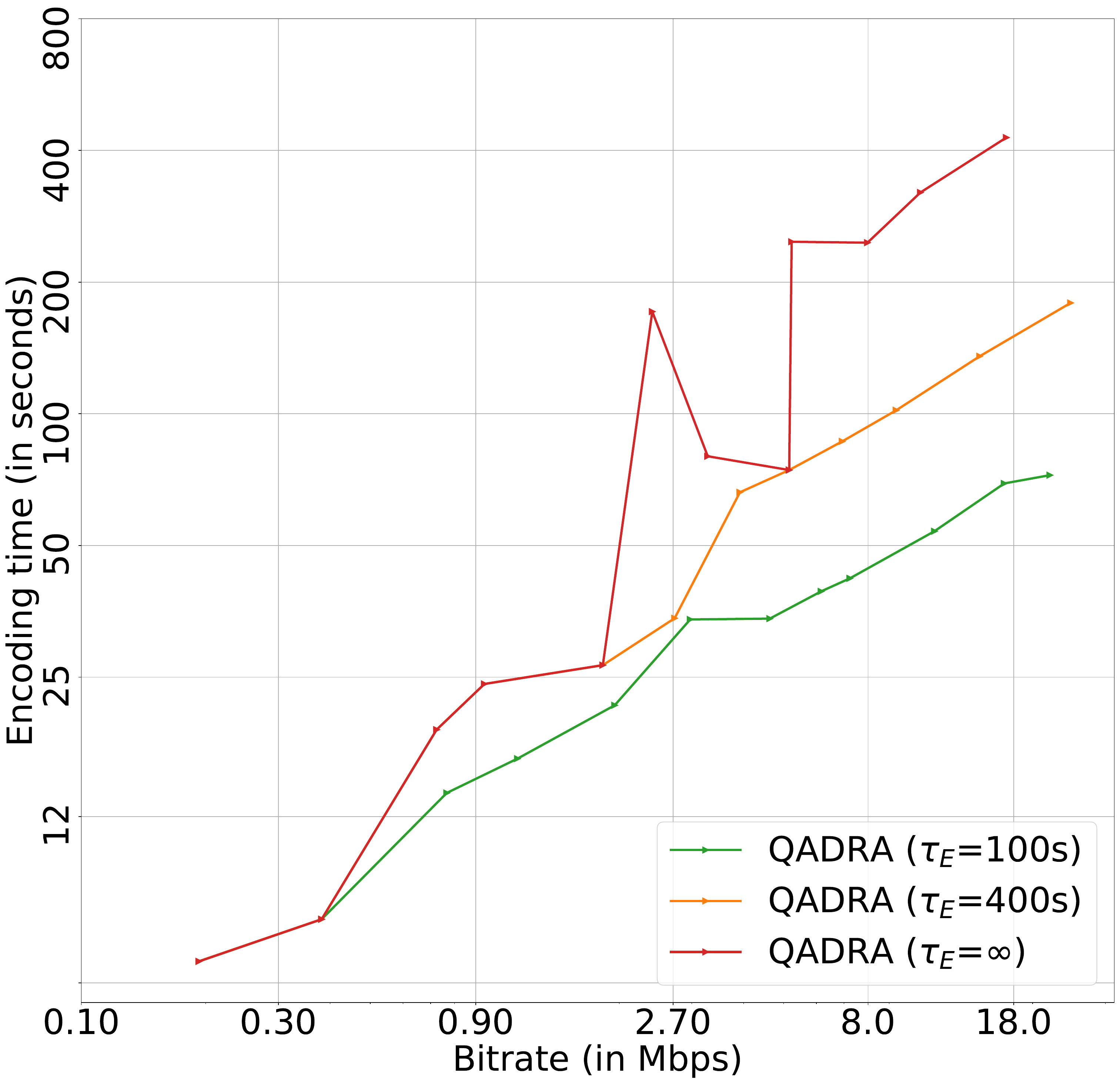}
    \caption{\textit{0001}}
\end{subfigure}
\begin{subfigure}{0.223\textwidth}
    \centering    
    \includegraphics[clip,width=0.9\textwidth]{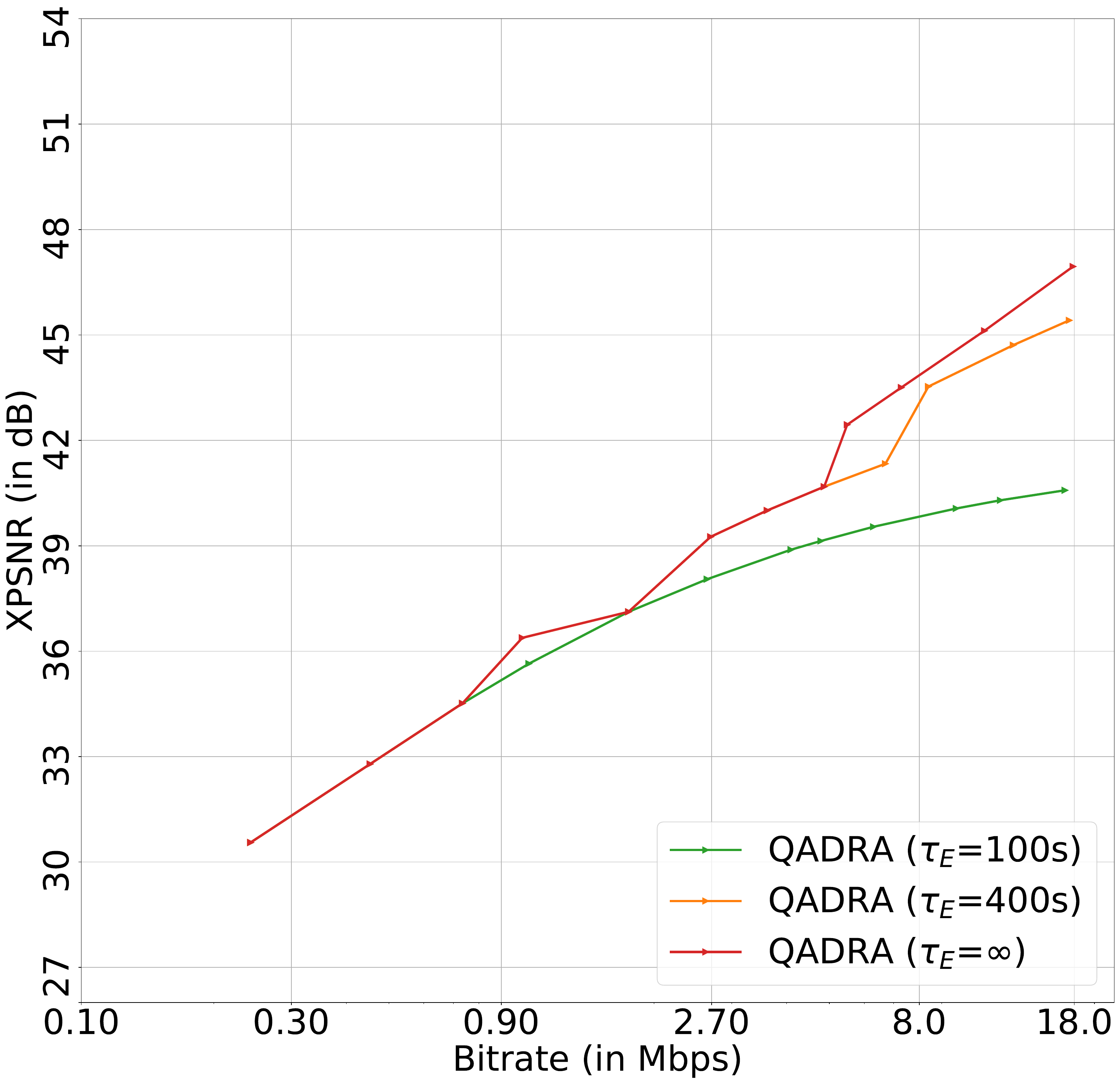}  
    \includegraphics[clip,width=0.9\textwidth]{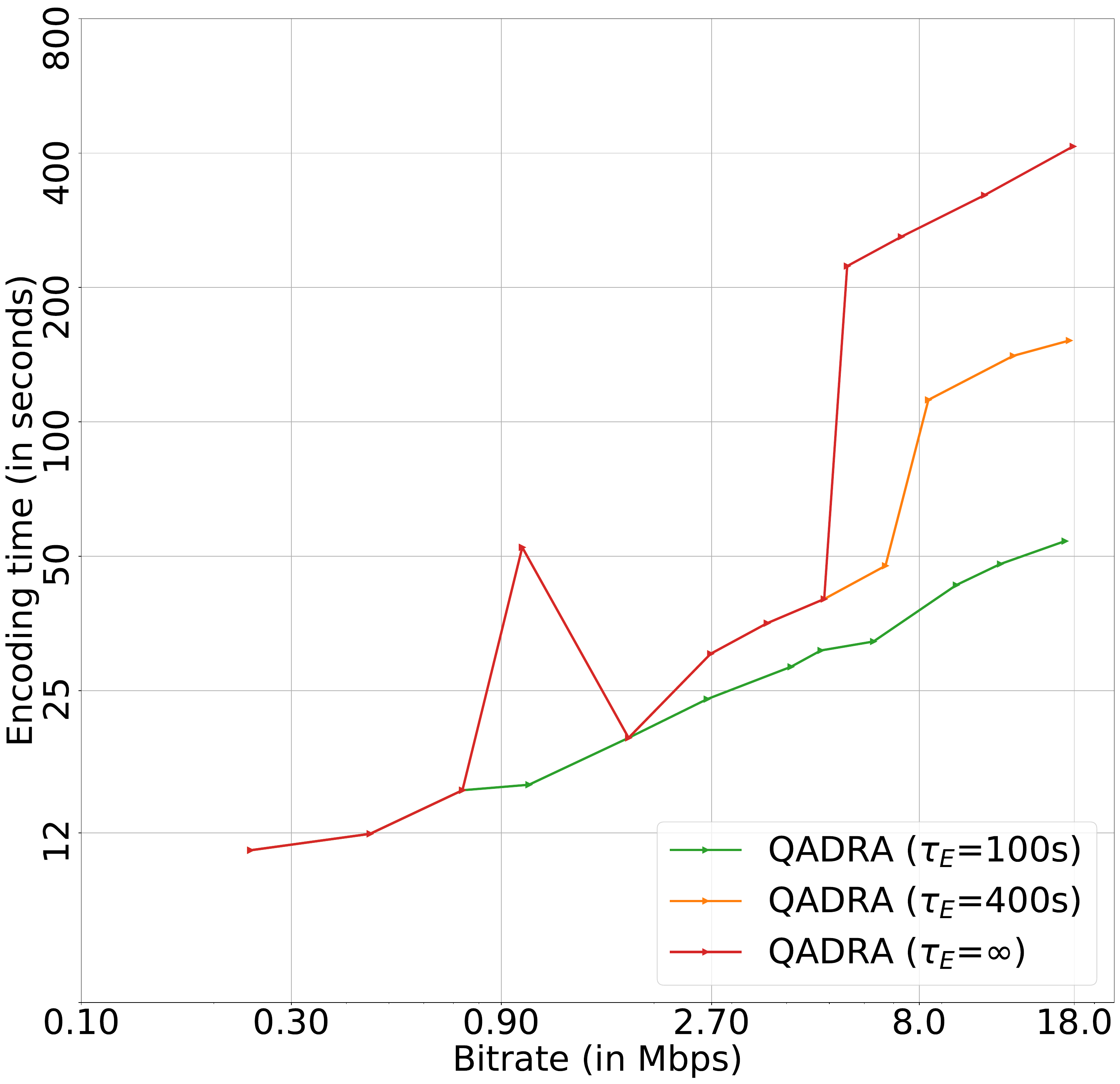}
    \caption{\textit{0334}}
\end{subfigure}
\begin{subfigure}{0.223\textwidth}
    \centering    
    \includegraphics[clip,width=0.9\textwidth]{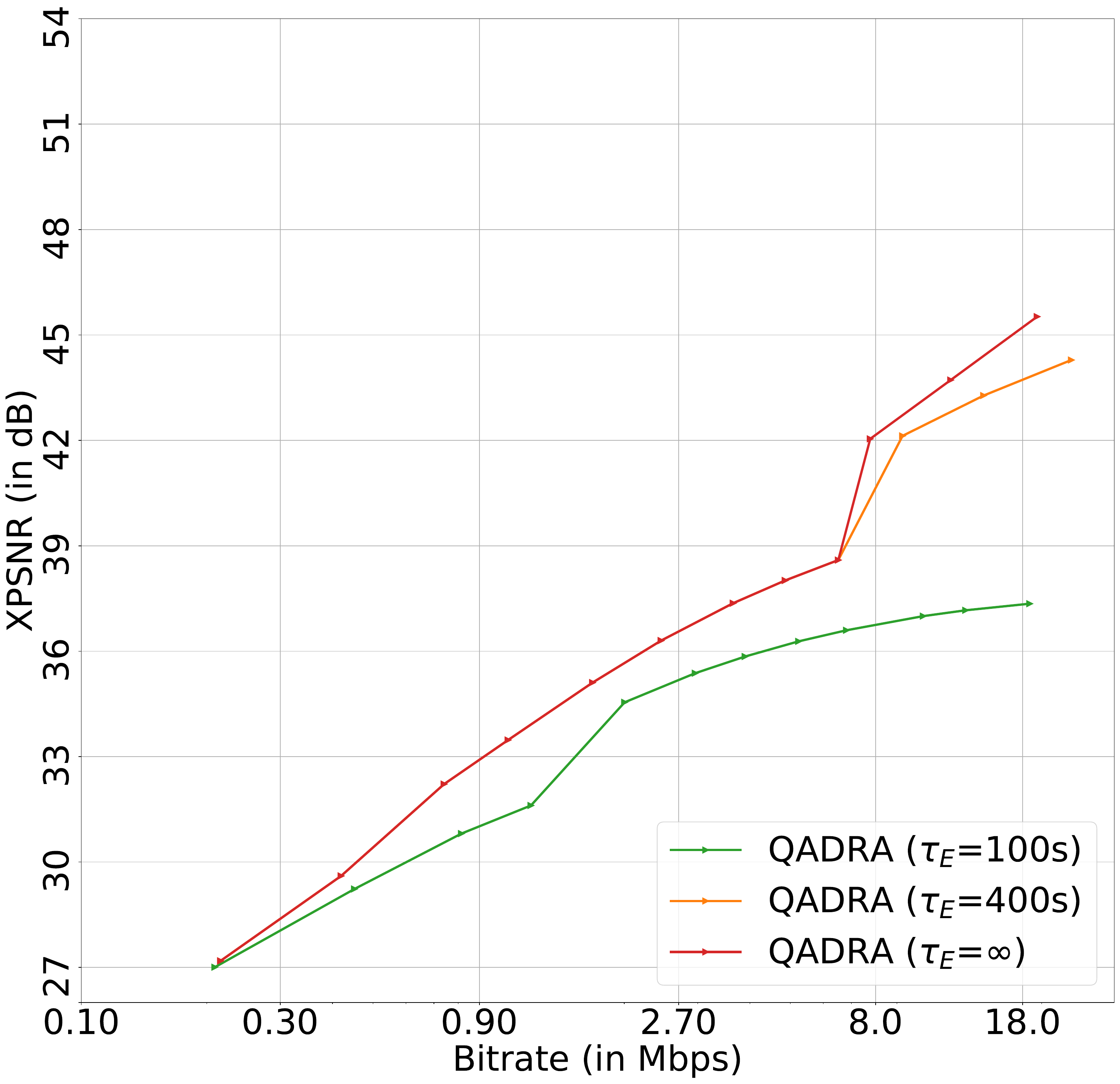}  
    \includegraphics[clip,width=0.9\textwidth]{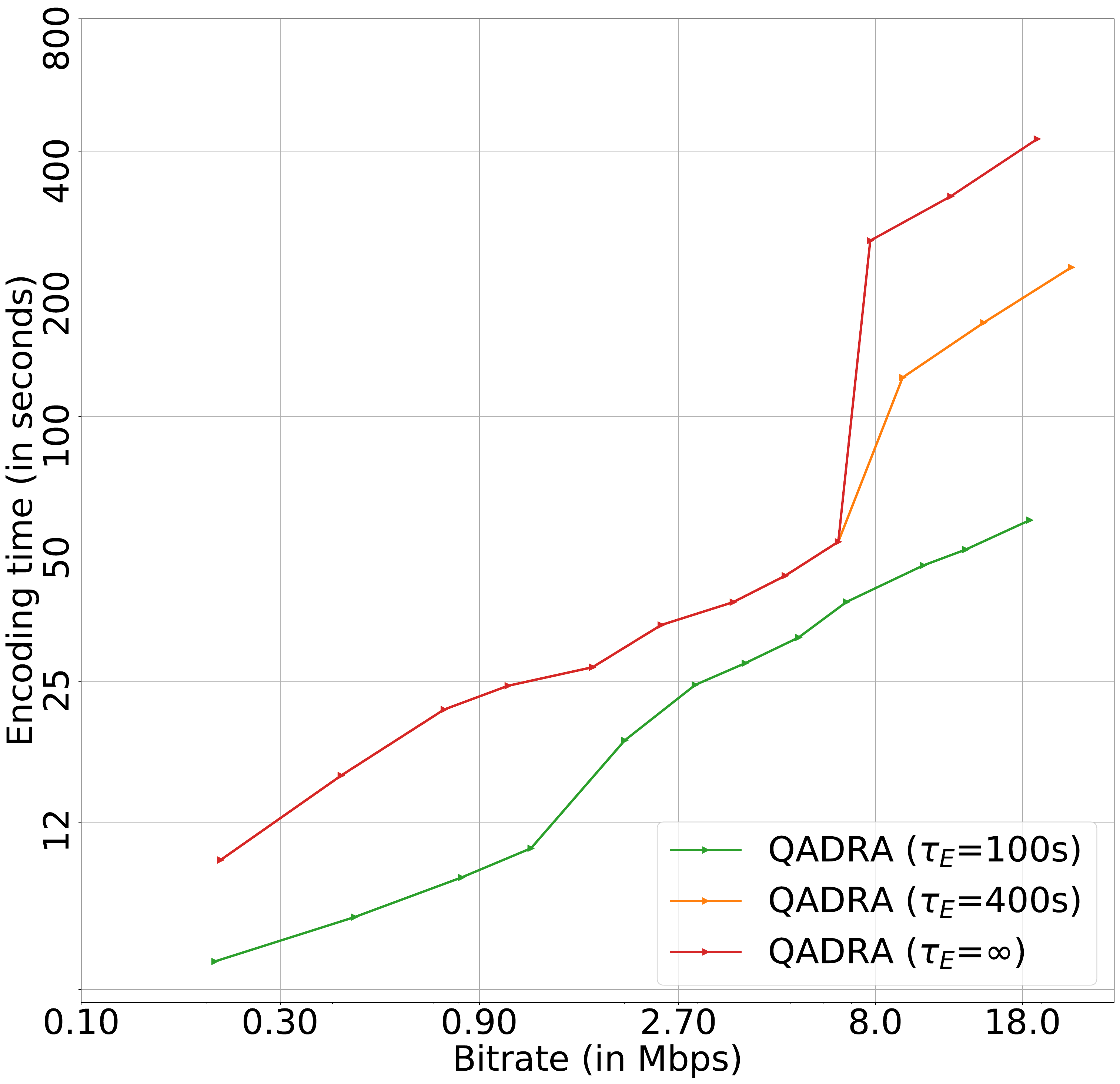}
    \caption{\textit{0412}}
\end{subfigure}
\begin{subfigure}{0.223\textwidth}
    \centering
    \includegraphics[clip,width=0.93\textwidth]{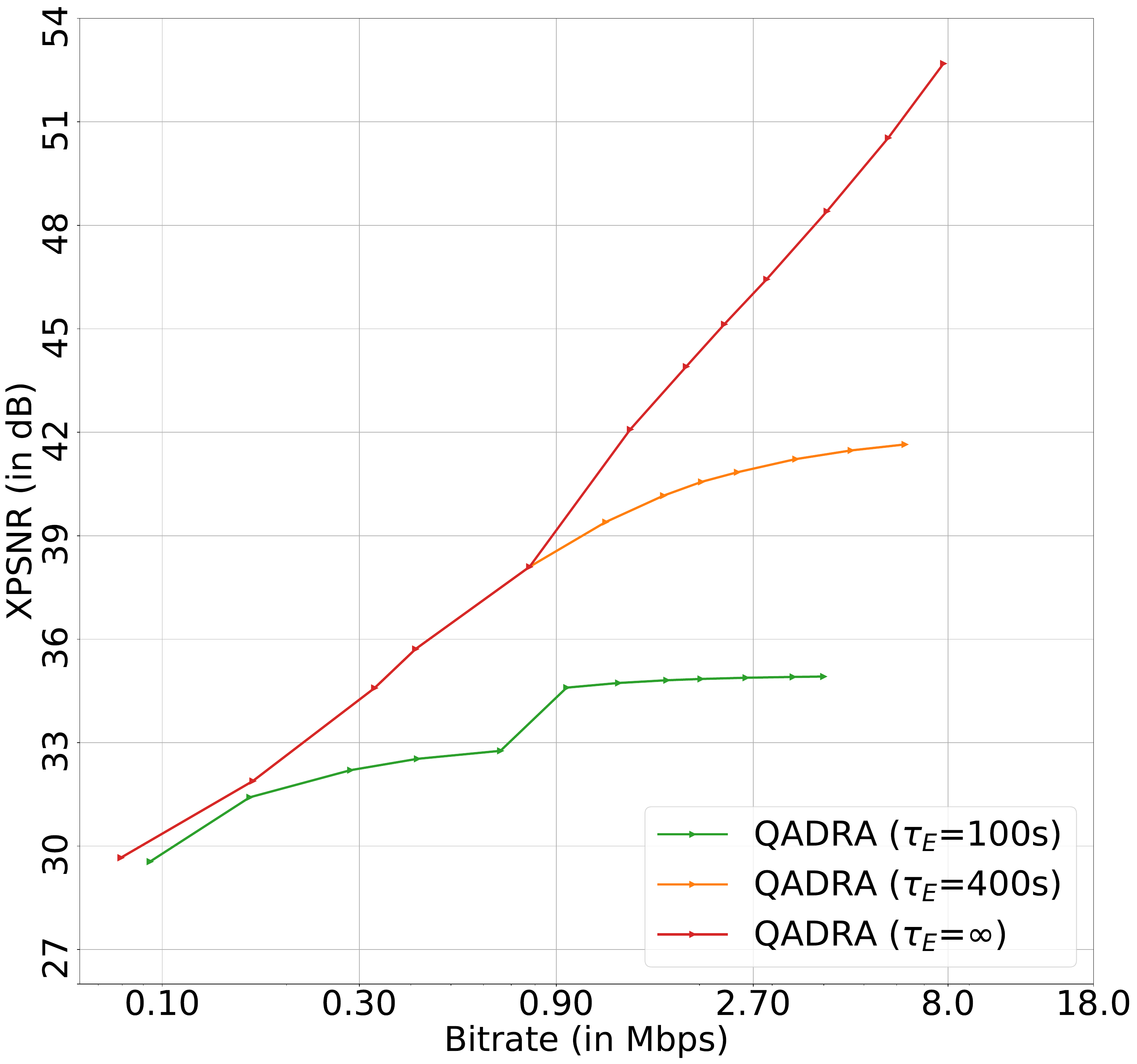}  
    \includegraphics[clip,width=0.94\textwidth]{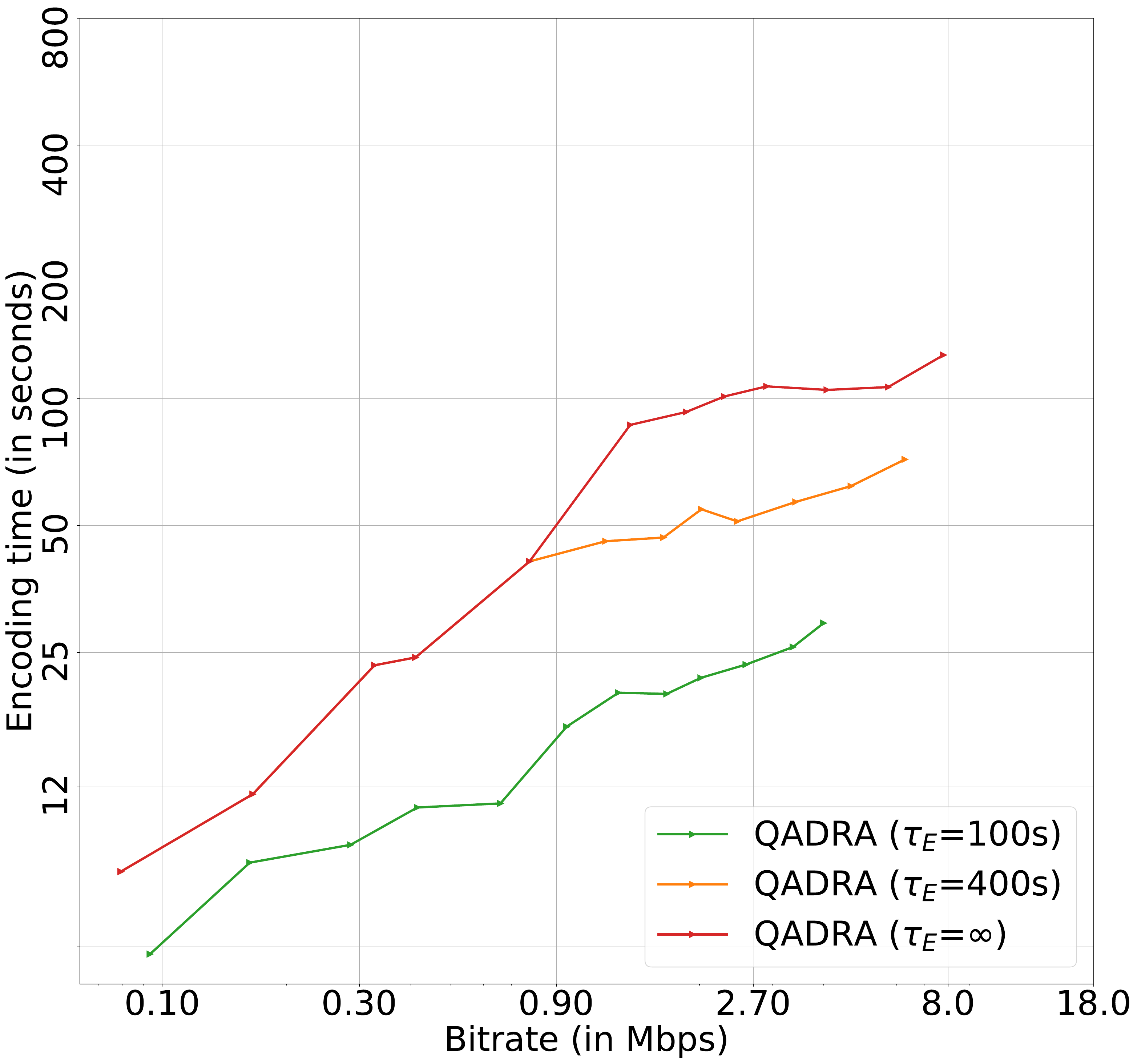}
    \caption{\textit{0744}}
\end{subfigure}
\caption{RD curves, and encoding times of representative video sequences (segments) using \scheme ($r_{\text{max}}=2160$).}
\label{fig:rd_res}
\end{figure}

\section{Evaluation results}
\label{sec:results}
The experimental parameters used to evaluate \scheme are listed in Table~\ref{tab:exp_par}. The VVenC encoding uses the predicted bitrate-resolution-QP configurations for a given input video segment. $\hat{b}_{t}$ is considered the upper bound of bitrate variability, and $\hat{q}_t$ is the QP used for encoding. In VVenC, the QP is specified using the \texttt{qp} option, while the \texttt{maxrate} (easy mode) or \texttt{MaxBitrate} (expert mode) option is used to specify the upper bound of bitrate variability.

\textbf{\textit{Prediction latency and accuracy: }} 
The time to predict the resolution-QP for each representation is \SI{5}{\milli\second}. The accuracy of the encoding time, QP, and XPSNR prediction models are analyzed in terms of mean absolute error (MAE). The average MAE is \SI{56.69}{\second}, 1.32, and \SI{0.16}{\decibel}, respectively. The average standard deviation is \SI{83.98}{\second}, 1.96, and \SI{0.22}{\decibel}, respectively.

\textbf{\textit{Resolution prediction: }} 
The encoding resolution predictions of \scheme are analyzed. Figure~\ref{fig:res_res} shows that \scheme ($\tau_{E}=\infty$) generally yields the highest resolutions for a given target bitrate compared to \textit{Default} and other encodings. The selected encoding resolution for a given target bitrate decreases as $\tau_{\text{E}}$ decreases. If the target latency constraint in \scheme is eliminated, \ie $\tau_{\text{E}}=\infty$, resolutions yielding the highest XPSNR are selected. Notably, in scenarios where encoding time constraints become more stringent, higher bitrate representations might be omitted in \scheme due to limitations in encoding these representations within the allocated time budget, as observed in Figure~\ref{fig:rd_res}.

\textbf{\textit{Rate-distortion performance: }} 
Figure~\ref{fig:rd_res} shows the RD curves of the representative video segments in the test dataset. It is observed that the RD curve of \scheme ($\tau_{\text{E}}=\infty$) is consistently higher than \textit{Default} and other encodings. This means that, for any given target bitrate, \scheme ($\tau_{\text{E}}=\infty$) maintains a higher level of visual quality as measured by XPSNR. Consequently, viewers can enjoy a visually pleasing and immersive experience with reduced artifacts, such as blocking or blurring, at the same bitrate.

\textbf{\textit{Latency and energy consumption performance: }} 
As shown in Figure~\ref{fig:rd_res}, \scheme ($\tau_{\text{E}}=\infty$) yields the longest encoding time due to higher encoding resolutions optimized for maximum XPSNR. This significantly increased encoding time may impact real-time or low-latency applications. Encoding typically utilizes the processing units (\eg CPU or GPU) intensively. These processing units operate at a relatively constant power level during encoding. Therefore, the power consumed over time remains reasonably consistent, contributing to the linear relationship between encoding time and energy consumption. Hence, we assume that the encoding time savings directly translates to the encoding energy consumption reduction. Since the encodings are assumed to be carried out concurrently, the total encoding time for each segment ($\tau_{\text{E}}$) is determined to be the highest encoding time yielded among the bitrate ladder representations~\cite{emes_ref}. Table~\ref{tab:res_cons} shows the average encoding time for each segment ($\overline{\tau_{\text{E}}}$) using the considered encoding schemes. It is observed that the encoding and decoding times of representations of video segments decrease as $\tau_{\text{E}}$ decreases.

\begin{table}[t]
\caption{Average encoding and decoding time results of \scheme compared to the \textit{Default} bitrate ladder encoding~\cite{apple_inc_http_nodate}.}
\centering
\resizebox{0.52\columnwidth}{!}{
\begin{tabular}{l|c||c|c|c}
\specialrule{.12em}{.05em}{.05em}
\specialrule{.12em}{.05em}{.05em}
$r_{\text{max}}$ & $\tau_{\text{E}}$ & $\Delta T_{\text{E}} \approx \Delta E$ & $\overline{\tau_{\text{E}}}$ &  $\Delta T_{\text{D}}$ \\
& [s] & [\%] & [s] & [\%] \\
\specialrule{.12em}{.05em}{.05em}
\specialrule{.12em}{.05em}{.05em}
\multirow{5}{*}{720} & 100       &  -66.57   &  59.38   & -71.77 \\
 & 200       &  -49.94   &  95.40   & -57.66 \\
 & 400       &  -49.51   &  96.40   & -57.26 \\
 & 800       &  -49.51   &  96.40    & -57.26 \\
 & $\infty$  &  -49.51   &  96.40    & -57.26 \\
\hline
\multirow{5}{*}{1080} & 100      &  -66.55   &  59.38    &  -71.74\\
 & 200      &  -41.45   &  112.08   & -47.43 \\
 & 400      &  -19.53   &  180.16   & -26.77 \\
 & 800      &  -19.39   &  181.00   & -26.70 \\
 & $\infty$ &  -19.39   &  181.00   & -26.70 \\
\hline
\multirow{5}{*}{2160} & 100  &    -66.55    & 59.38 & -71.74 \\
 & 200 &     -41.30    & 112.08 & -47.13 \\
 & 400 &     -9.15    & 198.11 & -10.11 \\
 & 800 &     31.22    & 318.05 & 45.50\\
 & $\infty$ & 41.56   & 364.95 & 57.25 \\
\specialrule{.12em}{.05em}{.05em}
\specialrule{.12em}{.05em}{.05em}
\end{tabular}
}
\label{tab:res_cons}
\end{table}

\section{Conclusions and future directions}
\label{sec:conclusion}
This paper implemented a quality-aware dynamic resolution adaptation (\scheme) framework for adaptive streaming applications. \scheme implements an optimized resolution and QP prediction, which uses XGboost-based models to estimate bitrate-resolution-QP triples for a given video segment based on spatiotemporal characteristics. Furthermore, a JND-aware representation elimination algorithm is also implemented, removing the bitrate ladder's perceptual redundancy.

One promising avenue for future research is exploring advanced machine-learning models to enhance prediction accuracy. Investigating novel features and metrics that better capture the relationship between encoding time and optimal resolutions might also be a promising avenue. Moreover, delving into collaborative frameworks or distributed algorithms for efficient encoding resolution selection across multiple streaming nodes could be another area of exploration.
\balance
\newpage
\bibliographystyle{IEEEtran}
\bibliography{references}
\balance
\end{document}